\documentstyle[aps,preprint,tighten]{revtex}
\def\d{\partial}
\def\p{{\vec p}}
\def\mD{m_{\rm D}}
\def\<{\langle}
\def\>{\rangle}

\firstfigfalse
\begin{document}

\title{``Bottom-up'' thermalization in heavy ion collisions}

\author{R.~Baier$^a$, A.H.~Mueller$^b$, D.~Schiff$^c$, and D.T.~Son$^{b,d}$}

\address{
$^a$ Fakult\"at f\"ur Physik, Universit\"at Bielefeld,
D-33501 Bielefeld, Germany\\
$^b$ Physics Department, Columbia University, New York, NY 10027, USA\\
$^c$ LPT, Universit\'e Paris-Sud, B\^atiment 210, F-91405 Orsay, France\\
$^d$ RIKEN-BNL Research Center, Brookhaven National Laboratory, Upton,
NY 11973, USA}

\maketitle

\begin{abstract}
We describe how thermalization occurs in heavy ion collisions in the
framework of perturbative QCD.  When the saturation scale $Q_s$ is
large compared to $\Lambda_{\rm QCD}$, thermalization takes place
during a time of order $\alpha^{-13/5}Q_s^{-1}$ and the maximal
temperature achieved is $\alpha^{2/5}Q_s$.
\end{abstract}

\section{Introduction}

It is possible that at RHIC, for the first time, heavy ion collisions
occur at energies high enough to be described by perturbative QCD.  At
the Large Hadron Collider (LHC) perturbative QCD is expected to work
even better.  At these energies we will assume that immediately after
the collision the initial distribution of gluons is given by the
saturation scenario \cite{GLR,Blaizot,McLVen,JKMW,Eskola,Mueller}.
Thus the relevant hard scale is the saturation scale $Q_s$, estimated
to be 1 GeV at RHIC and 2-3 GeV at LHC.

The single most important question in the physics of heavy ion
collisions is thermalization.  The conventional argument in favor of
thermalization is that at larger collision energy, more gluons are
freed in the first moment after the collision, and these gluons
collide more frequently with each other.  However, the distribution of
these gluons is initially very far from thermal equilibrium.  In
addition, the strong coupling constant decreases at high energies,
making thermalization harder to achieve.  Whether the system has
enough time to equilibrate before falling apart is thus a delicate
question requiring detailed consideration of different physical
processes.

In this paper we show that, in the limit $Q_s\gg\Lambda_{\rm QCD}$
corresponding to very large nuclei and/or very high collision energy,
thermalization occurs relatively fast while the system is still
undergoing one-dimensional expansion.  The unexpected feature of our
analysis is the way thermalization occurs.  During the first period of
time the most important process is the emission of soft gluons which
overwhelm, in terms of number, the primary hard gluons at time
$\tau\sim \alpha^{-5/2} Q_s^{-1}$.  These soft gluons then quickly
equilibrate and form a thermal bath, which initially carries only a
small fraction of the total energy.  The thermal bath then draws
energy from the hard gluons.  Full thermalization is achieved when the
primary hard gluons have lost all their energy.  Parametrically, this
happens at $\tau \sim \alpha^{-13/5}Q_s^{-1}$, at which time the
temperature of the system achieves the maximal value of
$\alpha^{2/5}Q_s$.  Surprisingly, the time dependence of the
temperature of the soft sector in the (admittedly narrow) region
$\alpha^{-5/2} \ll Q_s\tau \ll \alpha^{-13/5}$ can be found almost
analytically, and the result depends only on the total number of
primary hard gluons.  It is important to emphasize here that we only
provide parametric qualitative and quantitative estimates based on the
assumption that always $\alpha\ll1$.  A more realistic study is
required in order to figure out the numerical coefficients of the
given estimates.

This picture of ``bottom up'' thermalization is different from that
considered in a previous work by one of us \cite{elastic}
(see also \cite{elastic2})
which did
not take into account particle production, and also of Refs.\
\cite{Biro,Shuryak,Wong,Elliott,GeigerMuller,Geiger}.
In \cite{Wong} the importance of inelastic processes, even for kinetic
equilibration, has been observed when using the relaxation time
approximation for the collision term and with initial conditions
different from the ones considered here.
Although the analysis requires a small coupling, one can hope that
many qualitative features of this picture are already present in heavy
ion collisions at LHC or even RHIC energies.  At the very least, the
finding gives us confidence that thermalization always occurs in heavy
ion collisions at sufficiently high energies.
Our main emphasis here is the thermalization that begins with the softer
momentum modes.  Still more detailed investigations, including more
specific realistic initial conditions, are required for treating the
high momentum tails of the distributions which are expected to require a
longer thermalization time, but which are important for plasma signatures
sensitive to momentum scales larger than $Q_s$.

\section{Qualitative description of early times}

We will be interested only in the central rapidity region of central
collisions.  In this region one can assume boost invariance: all
physical quantities depend only on the proper time
$\tau=\sqrt{t^2-z^2}$, but not on the rapidity
$\eta={1\over2}\ln{t+z\over t-z}$, where $z$ is the direction of
collision.  For large nuclei, the dependence on the transverse ($x$
and $y$) coordinates can also be neglected, which is equivalent to
assuming the medium to be infinite in the transverse directions.  With
these simplifications, all physical quantities depend only on the
single coordinate $\tau$.
The evolution proceeds through several regimes, with $Q_s\tau \sim
\alpha^{-3/2}$, $\alpha^{-5/2}$, and $\alpha^{-13/5}$ marking the
borders between neighboring time periods.

We will describe qualitatively the evolution of the system up to
$Q_s\tau\sim\alpha^{-5/2}$.  In all subsequent estimates, we will
assume $\ln{1\over\alpha}\sim1$ and keep track only of powers of
$\alpha$ itself.

\subsection{Very early time, $1 \ll Q_s\tau \ll \alpha^{-3/2}$}

At the earliest time, $\tau\sim Q_s^{-1}$, gluons are freed from the
nuclei.  These gluons have typical momentum of order $Q_s$ and
occupation number of order $1/\alpha$.  This is the basic assumption
of the rest of this paper.  Due to the large occupation number these
gluons interact so strongly that it is more appropriate to describe
them as a nonlinear gluon field rather than a collection of particles.
Only when $Q_s\tau$ becomes larger than 1, the classical field becomes
almost linear, and one can start to describe the gluons as particles
on mass shell with a well-defined distribution,
\begin{equation}
   {dN\over dy\, dk_\perp^2} = {1\over\alpha}
   f\biggl( {k_\perp\over Q_s}\biggr) \, .
\end{equation}
The precise form of $f$ depends on the details of the nonperturbative
physics at the scale $\tau\sim Q_s^{-1}$ and is not a subject of the
present paper.  A promising method to find $f$ is by simulation of the
classical gluon field \cite{KrasVen}.  We will call the particles
produced during these first periods the hard gluons, since later
gluons with smaller momenta will be produced.

The density of hard gluons
\cite{GLR,Blaizot,McLVen,JKMW,Eskola,Mueller} decreases with time due
to the one-dimensional expansion,
\begin{equation}
   N_h \sim {Q_s^3 \over \alpha (Q_s\tau)} \, .
   \label{Nh}
\end{equation}
If there was no interaction, the occupation number would remain of
order $1/\alpha$ due to the Liouville theorem.  There is no
contradiction with Eq.\ (\ref{Nh}), since the typical longitudinal
momentum of hard gluons also becomes smaller ($p_z\sim1/\tau$) because
gluons with larger longitudinal momentum escape from the spatial
region under consideration during a time of order $\tau$.

In reality, gluons interact by elastic scattering.  Most of the
scatterings are small angle, with exchange momentum $q\ll Q_s$.  The
effect is the broadening of the distribution along the $p_z$
direction, thus lowering the typical occupation number.  The lowest
possible momentum exchange is the Debye mass, which is determined by
\cite{Biro:1992ix,elastic2,Nayak}
\begin{equation}
   \mD^2 \sim \alpha\int\!d^3p\, {f_h(p) \over p}
   \sim {\alpha N_h\over Q_s} \sim {Q_s^2\over Q_s\tau} \, .
   \label{mDearly}
\end{equation}
If one assumes that $\mD\ll p_z$ (this condition will be verified {\em
a posteriori}), most collisions do not take particles away from the
momentum region where the occupation number is large ($k_\perp\sim
Q_s$, $k_z\sim p_z$).  The frequency of collisions that a typical
particle encounters is enhanced by the Bose factor,
\begin{equation}
   {dN_{\rm col}\over d\tau} \sim \sigma N_h (1+f_h) \sim
   {\alpha N_h \over \mD^2 p_z \tau} \, ,
\end{equation}
where $\sigma\sim \alpha^2 \mD^{-2}$ is the cross section, and
$f_h=N_h/(Q_s^2p_z)$ is the typical occupation number, which is
assumed to be large.  These random collisions increase the
longitudinal momentum of gluons, which is typically
\begin{equation}
   p_z^2 \sim N_{\rm col} \mD^2 \sim {\alpha N_h \over p_z} \, ,
\end{equation}
which implies
\begin{equation}
   p_z \sim (\alpha N_h)^{1/3} \sim {Q_s \over (Q_s\tau)^{1/3}} \, .
   \label{pz}
\end{equation}
This relation may be derived from \cite{elastic}, taking the Bose
enhancement into account, which amounts to replace the average
$p_z^2\sim\alpha Q_s^2$ by $p_z^2\sim\alpha_s Q_s^2(1+f_h)\sim
\alpha N_h/p_z$.
>From Eqs.\ (\ref{mDearly}) and (\ref{pz}) we see that $\mD\ll p_z$, as
assumed.  The typical occupation number $f_h \sim \alpha^{-1}
(Q_s\tau)^{-2/3}$ is large until $Q_s\tau\sim\alpha^{-3/2}$.

Beside elastic scatterings there are also inelastic scatterings in
which gluons are produced.  As we will see, the most important
produced gluons are those with smallest energies.  In principle gluons
with energy as low as $\mD$ can be produced.  However, once produced,
the momentum of these gluons is pushed up by multiple elastic
scatterings with hard gluons.
As a consequence of (\ref{pz})
the smallest
momentum of soft gluons is of order $p_z$: $k_s \sim p_z \sim
Q_s/(Q_s\tau)^{1/3}$.  The number of soft gluons with momentum $k_s$
at time $\tau$ that are produced at this moment
is estimated from the Bethe-Heitler formula to be \cite{Gunion}
\begin{equation}
   N_s \sim \tau {\d N_s \over\d\tau} \sim
   \tau\int\!d^3p\,f(p){dI^{\rm BH}\over dt}(1+f_h)^2 \sim
   \tau {\alpha^3\over \mD^2} N_h^2 (1+f_h)^2 \sim
   {Q_s^3\over\alpha(Q_s\tau)^{4/3}} \, .
\end{equation}
The time interval around the moment $\tau$ is taken to be of order
$\tau$.  Once these soft gluons are produced they will remain, and
their density is decreasing as $1/\tau$.  One notices that
$N_s/k_s\sim N_h/Q_s$, so the Debye mass receives {\em equal}
contributions from hard and soft gluons,
\begin{equation}
   \mD^2 \sim {\alpha N_h\over Q_s} + {\alpha N_s \over k_s} \, ,
\end{equation}
but the estimate of Eq.\ (\ref{mDearly}) is still valid
parametrically.

\subsection{Setting up the stage for thermalization:
$\alpha^{-3/2} \ll Q_s\tau \ll \alpha^{-5/2}$}

Beginning from $Q_s\tau \sim \alpha^{-3/2}$, the occupation number of
hard gluons drops below 1, and the estimates of the previous section
need to be revised.  We will see that when $\alpha^{-3/2} \ll Q_s\tau
\ll \alpha^{-5/2}$ soft gluons contribute negligibly to the total
number of gluons but give most of the Debye screening.  In other
words,
\begin{eqnarray}
   N_s & \ll & N_h \, , \nonumber\\
   \mD^2 & \sim & {\alpha N_s \over k_s} \, .
   \label{mDmed}
\end{eqnarray}
First let us estimate $k_s$, which is the typical momentum of soft
gluons.  When $f_h\ll 1$, we have
\begin{equation}
   k_s^2 \sim N_{\rm col} \mD^2 \sim
   \tau \sigma N_h \mD^2 \sim \alpha Q_s^2 \, ,
\end{equation}
which is now a constant scale. The number of soft gluons that have
been produced at time $\tau$ is
\begin{equation}
   N_s \sim \tau {\alpha^3\over \mD^2} N_h^2 \sim
   {\alpha Q_s^4\over \mD^2 \tau} \, .
   \label{Nsmed}
\end{equation}
>From Eqs.\ (\ref{mDmed}) and (\ref{Nsmed}), one finds
\begin{eqnarray*}
   N_s & \sim & {\alpha^{1/4} Q_s^3 \over (Q_s\tau)^{1/2}} \, , \\
   \mD & \sim & {\alpha^{3/8}Q_s \over (Q_s\tau)^{1/4}} \, .
\end{eqnarray*}
For the soft gluons to give the dominant contribution to Debye
screening one needs $N_s/k_s\gg N_h/Q_s$, which requires $Q_s\tau \gg
\alpha^{-3/2}$.  The number of soft gluons become comparable to that
of hard ones at $Q_s\tau \sim \alpha^{-5/2}$.  Therefore all estimates
in this section are valid in the interval $\alpha^{-3/2} \ll Q_s\tau
\ll \alpha^{-5/2}$.

In-medium emission of gluons is suppressed by the
Landau-Pomeranchuk-Migdal (LPM) effect. The latter is operative at
scales larger than $k_{\rm LPM}$ determined by $k_{\rm
LPM}=\mD^2(N_h\sigma)^{-1}$ \cite{BDMS,BSZ}, or
\begin{equation}
   k_{\rm LPM} \sim {\mD^4 \over \alpha^2 N_h} \sim \alpha^{1/2} Q_s \, ,
\end{equation}
which is parametrically of the same order as the scale $k_s$.  The LPM
effect suppresses the gluon production compared to the Bethe-Heitler
rate at scales larger than $k_s$, but does not affect the rate of
producing particles in the interval between $\mD$ and $k_s$.
Therefore all previous estimates based on the Bethe-Heitler formula
remain intact.

\section{Thermalization of the soft sector: $Q_{\lowercase{s}}\tau \gg
\alpha^{-5/2}$}

\subsection{Qualitative description}

After $Q_s\tau\sim\alpha^{-5/2}$ most gluons are soft, $N_s\gg N_h$.
We will see that the soft gluons collide very frequently with each
other, and achieve thermal equilibration amongst themselves.  The soft
sector is characterized by the temperature $T$, which is a function of
time.  The system as a whole is still not in thermal equilibrium,
since most of the energy is carried by a small number of hard gluons.
These few gluons collide with the soft gluons of thermal bath and
constantly loose energy to the latter.

A hard gluon with energy of order $Q_s$ looses energy to the bath by
the following mechanism.  First it emits a particle with a softer
momentum $k_{\rm br}$, which, during a time comparable to $\tau$,
splits into two gluons with comparable momenta (hard branching).  The
products of this branching quickly cascade further, giving all their
energy to the thermal bath.

As we will verify, $k_{\rm br}$ lies in the region where the emission
rate is LPM-suppressed.
One can estimate the time of emission, $t_{\rm br}$, for a gluon having
momentum $k_{\rm br}$ as follows: $1/t_{\rm br}\sim\alpha/t_{\rm f}$,
with the formation time $t_{\rm f}$ given by $t_{\rm f}\sim
k_{\rm br}/k_t^2$.  The gluon picks up a momentum $k_t$, transverse to
its direction of motion, given by $k_t^2\sim\mD^2 t_{\rm f}/\lambda$.
Using $\lambda^{-1}\sim N_s\sigma\sim N_s\alpha^2/\mD^2$ one obtains
(see also Eq.\
(\ref{dI./dx}) below),
\begin{equation}
   {1\over t_{\rm br}} \sim {\alpha^2 N_s^{1/2} \over
   k_{\rm br}^{1/2}} \, .
   \label{LPM}
\end{equation}
Equating the branching time $t_{\rm br}$ with $\tau$, and using
$N_s\sim T^3$, where $T$ is the temperature of the soft thermal bath,
we find that $k_{\rm br}\sim \alpha^4 T^3 \tau^2$.  The time
dependence of the temperature $T$ is still to be found.

The number of $k_{\rm br}$-gluons produced per unit time per unit
volume is
\begin{equation}
   {d N(k_{\rm br})\over d \tau} \sim
   {N_h \over t_{\rm br}} \sim
   {\alpha^2 N_s^{1/2} N_h \over k_{\rm br}^{1/2}} \sim
   {Q_s^2 \over \alpha \tau^2} \, .
\end{equation}
Subsequently, the rate of energy flow from the hard gluons to the soft
thermal bath is
\begin{equation}
   k_{\rm br} {d N(k_{\rm br}) \over d\tau} \sim \alpha^3 Q_s^2 T^3 \, .
\end{equation}
This energy flow increases the energy in the thermal bath, and thus
must be proportional to $d(T^4)/ d\tau$.  Therefore one finds
\begin{equation}
   T \sim \alpha^3 Q_s^2 \tau \, .
\end{equation}
The temperature of the soft thermal bath increases linearly with time,
even when the system is expanding, due to the hard gluons which serve
as an energy source.  When the bath starts forming
($Q_s\tau\sim\alpha^{-5/2}$), its temperature is $\alpha^{1/2}Q_s$.
The relaxation time of the soft sector is of order $\tau_{\rm
rel}\sim(\alpha^2T)^{-1}\sim(\alpha^5Q_s^2\tau)^{-1}$.  When
$Q_s\tau\gg\alpha^{-5/2}$, $\tau_{\rm rel}\ll \tau$, which justifies
the assumption of thermal equilibration of the soft sector,
i.e.\ in the relaxation time approximation
$f_s\to f^{\rm eq} (1-\exp(-\tau/\tau_{\rm rel}))$.

The linear growth of $T$ terminates when the hard gluons loose all of
their energy.  This happens when $k_{\rm br}\sim Q_s$, or
$\tau\sim\alpha^{-13/5}Q_s^{-1}$, when the temperature achieves a
maximal value of order $\alpha^{2/5}Q_s$, which is larger than the
initial temperature only by a factor of $\alpha^{-1/10}$.
Subsequently the temperature decreases as $\tau^{-1/3}$
\cite{Bjorken}.

The border between the Bethe-Heitler and LPM regimes is at $k_{\rm
LPM} = \mD^4/(\alpha^2N_s)\sim T\sim \alpha^3Q_s^2\tau$.  On the other
hand $k_{\rm br}\sim\alpha^4T^3\tau^2\sim\alpha^{13}Q_s^6\tau^5$,
which is much larger than $k_{\rm LPM}$ when
$Q_s\tau\gg\alpha^{-5/2}$.  Thus the use of the LPM formula in Eq.\
(\ref{LPM}) is justified.

\subsection{Quantitative description}

The picture given above can made quantitative in the form of a
Boltzmann equation describing the kinetics of the branching process.
In this kinetic approach the evolution of hard modes follows the
equation:
\begin{equation}
   \biggl({\d \over \d\tau} - {p_z\over\tau}{\d \over \d p_z}\biggr)f(\p)
   = C_{\rm el} + C_{\rm prod} \, ,
   \label{Boltzmann}
\end{equation}
where $C_{\rm el}$ is the elastic collision integral (see Ref.\
\cite{elastic}) and
\begin{eqnarray}
   C_{\rm prod} &=& \int\limits_0^1\!dx\, {d^2 I \over dx\, dt}
   \biggl\{ {1 \over x^{5/2}}
   \biggl[ f({\p\over x})(1{+}f(\p)) (1{+}f({\p(1{-}x)\over x})) -
   f(\p)f({\p(1{-}x)\over x})(1{+}f({\p\over x}))\biggr] \nonumber\\
   & & -{1\over2} \biggl[f(\p)(1{+}f(\p x))(1{+}f(\p(1{-}x)))
     -f(\p x)f(\p(1{-}x))(1{+}f(\p))\biggr]
   \biggr\} \label{Cprod}
\end{eqnarray}
is the term describing the $2\to3$ and $3\to2$ processes.  The four
terms in the collision integral correspond to the diagrams (a), (c),
(b) and (d) in Fig.\ \ref{figure}.  The black blob in this Figure
represents multiple scatterings off individual gluons in the medium.
In Eq.\ (\ref{Cprod}) $d^2I/dxdt$ is the rate of a hard gluon with
momentum $\p\sim Q_s$ to split almost collinearly into two gluons with
momenta $\p x$ and $\p(1-x)$ while moving in a medium where most
particles carry much smaller momenta.  This rate is LPM-suppressed and
can be computed using the method of Ref.\ \cite{BDMS,BSZ,Zakharov},
for an infinite size medium, which yields
\begin{equation}
   {d^2I\over dxdt} = {\alpha^2 N^{1/2} \over p^{1/2}} h(x)
   \label{dI./dx}
\end{equation}
where
\begin{equation}
   h(x) = h_0 {(1-x+x^2)^{5/2}\over(x-x^2)^{3/2}}\, , \qquad
   h_0  =  {2\over\pi^{1/2}} {N_c^2\over(N_c^2-1)^{1/2}}
         \biggl(\ln{\<k_t^2\>\over \mD^2}\biggr)^{1/2} \label{h} , \,
\end{equation}
and
\begin{equation}
   N = 2(N_c^2-1)\int\!{d\p\over(2\pi)^3}f(\p)(1+f(\p)) \, .
\end{equation}
The $N$ in (21) represents the density of possible scatterers in 
the system.
$N$ is not quite the same as the density of gluons because of the extra factor
of $(1+f)$, a factor necessary to correctly give the interaction rate. From the
discussion given earlier we expect $N_s \gg N_h$ so long as $Q_s\tau >
\alpha^{-5/2}$ and so we expect $N$ to be dominated by the soft (thermalized)
particles in our present discussion.

$h(x)$ has a symmetry property that $h(x)=h(1-x)$, using which one can
check that the collision term (\ref{Cprod}) conserves energy,
$\int\!d\p\,pC_{\rm prod}(p)=0$.  One might wonder why $C_{\rm prod}$
is of order $\alpha^2$ instead of $\alpha^3$.  An intuitive derivation
for this dependence is already given just above Eq.\ (\ref{LPM}).
\begin{figure}

\setlength{\unitlength}{1mm}
\begin{picture}(150,65)(0,10)

\thicklines

\put(0,65){\makebox(0,0)[l]{$C_{\rm prod}=$}}

\put(25,65){\vector(1,0){8}}
\put(33,65){\line(1,0){7}}
\put(40,65){\vector(3,2){12}}
\put(40,65){\vector(3,-2){12}}
\put(40,65){\circle*{5}}
\put(22,65){\makebox(0,0)[r]{$\displaystyle{\p\over x}$}}
\put(55,73){\makebox(0,0)[l]{$\p$}}
\put(55,57){\makebox(0,0)[l]{$\displaystyle{\p(1{-}x)\over x}$}}
\put(40,50){\makebox(0,0){(a)}}

\put(75,65){\makebox(0,0){$-$}}
\put(85,65){\makebox(0,0){${\displaystyle{1\over2}}$}}

\put(100,65){\vector(1,0){8}}
\put(108,65){\line(1,0){7}}
\put(115,65){\vector(3,2){12}}
\put(115,65){\vector(3,-2){12}}
\put(115,65){\circle*{5}}
\put(97,65){\makebox(0,0)[r]{$\p$}}
\put(130,73){\makebox(0,0)[l]{$\p(1{-}x)$}}
\put(130,57){\makebox(0,0)[l]{$\p x$}}
\put(115,50){\makebox(0,0){(b)}}

\put(37,30){\vector(1,0){15}}
\put(25,38){\vector(3,-2){6}}
\put(31,34){\line(3,-2){6}}
\put(25,22){\vector(3,2){6}}
\put(31,26){\line(3,2){6}}
\put(37,30){\circle*{5}}
\put(22,38){\makebox(0,0)[r]{$\p$}}
\put(22,22){\makebox(0,0)[r]{$\displaystyle{\p(1{-}x)\over x}$}}
\put(55,30){\makebox(0,0)[l]{$\displaystyle{\p\over x}$}}
\put(37,15){\makebox(0,0){(c)}}
\put(0,30){\makebox(0,0)[l]{$-$}}

\put(75,30){\makebox(0,0){$+$}}
\put(85,30){\makebox(0,0){${\displaystyle{1\over2}}$}}

\put(112,30){\vector(1,0){15}}
\put(100,38){\vector(3,-2){6}}
\put(106,34){\line(3,-2){6}}
\put(100,22){\vector(3,2){6}}
\put(106,26){\line(3,2){6}}
\put(112,30){\circle*{5}}
\put(100,40){\makebox(0,0)[b]{$\p(1{-}x)$}}
\put(97,22){\makebox(0,0)[r]{$\p x$}}
\put(130,30){\makebox(0,0)[l]{$\p$}}
\put(112,15){\makebox(0,0){(d)}}

\end{picture}
\caption{The diagrammatic representation of $C_{\rm prod}$}
\label{figure}
\end{figure}
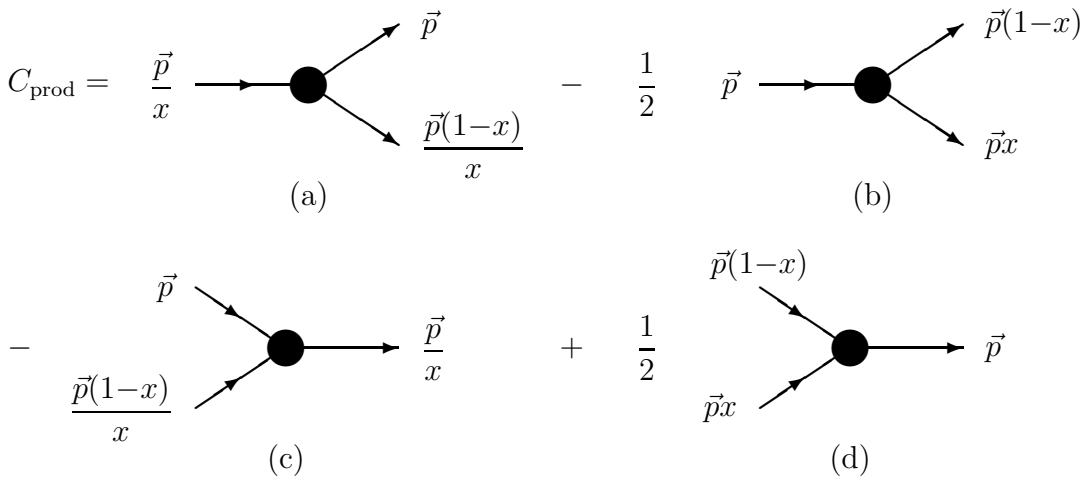\vspace{0.1mm}
A remark on the Boltzmann equation is in order.  In Ref.\
\cite{elastic} a Boltzmann equation without particle production was
considered.  The system does seem to approach kinetic equilibration
\cite{elastic,elastic2}, but during a relatively long time ($\sim
\exp(\alpha^{-1/2})Q_s^{-1}$).  This is because most of the elastic
scatterings are at small angle, and it takes a lot of small angle
scatterings to change the particle distribution considerably.  In more
technical terms, there is a cancellation between gain and loss terms
in the elastic collision integral.  The $2\to3$ process, on the other
hand, is asymmetric and does not have this type of cancellation, and
so is important although it is smaller than the elastic rate.
Parametrically higher-order processes like $2\to4$ can be neglected
since they are suppressed by an additional factor of $\alpha$ compared
to $2\to3$ while not adding any qualitatively new feature to the
evolution.
Whether in a realistic circumstance, say at RHIC, higher order processes
are negligible or not is beyond the scope of the present discussion
\cite{Shuryak}.

The elastic collision integral $C_{\rm el}$ is responsible for the
thermalization of the soft sector $p\sim T$, and for $p_z$ broadening
of the hard particles.  The inelastic term $C_{\rm prod}$ is
responsible for the energy flow from the hard to the soft sector.  In
principle, the products of branching of hard gluons have a small
transverse momentum, but this is negligible compared to the
subsequent broadening by elastic scattering.  Therefore, $C_{\rm
prod}$ is written as if the branching was exactly collinear.

For hard gluons which have small occupation number ($f\ll1$) Eq.\
(\ref{Boltzmann}) simplifies considerably.  If we introduce
\begin{equation}
   \epsilon({p_\perp}) = 2\pi p_\perp^2 \int\!dp_z\, f(\p)
\end{equation}
which is normalized so that the energy carried by hard gluons is
$\int\!dp_\perp\epsilon(p_\perp)$, then the kinetic equation for
$\epsilon(p_\perp)$ becomes
\begin{equation}
   {1\over \tau}{\d \over \d \tau}(\tau\epsilon(p_\perp)) =
   {\alpha^2 N^{1/2} \over p_\perp^{1/2}}
   \int\! dx\, h(x) \biggl[x^{1/2} \epsilon\biggl({p_\perp\over x}\biggr)
   - {1\over 2} \epsilon(p_\perp)\biggr] \, .
   \label{boltzeq}
\end{equation}
In the following we will write $p$ instead of $p_\perp$ for
simplicity.  Eq.\ (\ref{boltzeq}) does not describe the soft gluons,
for which $C_{\rm el}$ is essential.  Fortunately, these gluons are
fully equilibrated and one can characterize the soft sector by the
temperature $T$.  In particular,
\begin{equation}
   N = g_N T^3, \qquad
   g_N = 2(N_c^2-1){1\over 6} \, .
   \label{NT}
\end{equation}
The time evolution of $T$ depends on the amount of energy that flows
from the hard to the soft sector.  To find this quantity, let us
introduce an intermediate scale $p_0$, $T\ll p_0\ll Q_s$, and
integrate Eq.\ (\ref{boltzeq}) from $p_0$ to $\infty$.  One finds
\begin{equation}
   {1\over \tau}{\d \over \d \tau} \biggl(
   \tau\int\limits_{p_0}^\infty\!dp\,\epsilon(p)\biggr) =
   - \alpha^2 N^{1/2} \int\limits_0^1\! dx\, h(x) x
   \int\limits_{p_0}^{p_0/x}\! dp\, {\epsilon(p)\over p^{1/2}} \, .
   \label{flow}
\end{equation}
The left hand side of Eq.\ (\ref{flow}) has the meaning of the rate of
energy flow from above $p_0$ to below $p_0$.  This quantity must be a
constant independent of $p_0$ when $p_0\ll Q_s$.  This is possible only
when $\epsilon(p)$ has the following behavior at small $p$:
\begin{equation}
   \epsilon(p) = {\epsilon_1\over p^{1/2}}, \qquad T\ll p\ll Q_s \, .
   \label{esmallp}
\end{equation}
The right hand side of Eq.\ (\ref{flow}) is then $-bh_0\alpha^2
N^{1/2}\epsilon_1$, where
\begin{equation}
   b = \int\limits_0^1\!dx\, {(1-x+x^2)^{5/2}\over(x-x^2)^{3/2}}
       x\ln{1\over x} \approx
   4.96 \, .
\end{equation}
The equation that governs the evolution of the
temperature is then
\begin{equation}
   {1\over \tau^{4/3}} {\d \over \d \tau} (\tau^{4/3} \epsilon_s(T)) =
   bh_0\alpha^2 N^{1/2}\epsilon_1
   \label{Tt}
\end{equation}
because of the one-dimensional expansion \cite{Bjorken,Baym}
where $\epsilon_s(T)$ is the energy density of the soft gluons at
temperature $T$,
\begin{equation}
   \epsilon_s(T) = g_E T^4, \qquad g_E = 2(N_c^2-1) {\pi^2\over30} \, .
   \label{epsT}
\end{equation}
Eqs.\ (\ref{boltzeq}), (\ref{NT}), (\ref{Tt}), and (\ref{epsT}) are
the equations that govern the evolution after
$Q_s\tau\sim\alpha^{-5/2}$.  From Eq.\ (\ref{boltzeq}) one needs to
extract $\epsilon_1$, which enters Eq.\ (\ref{Tt}) that describes the
temperature $T$, which feeds back to Eq.\ (\ref{boltzeq}) through
$N$.

Let us show that this procedure can be performed analytically in the
regime $\alpha^{-5/2}\ll Q_s\tau\ll\alpha^{-13/5}$.  Eq.\
(\ref{boltzeq}) can be solved by iteration.  We will be looking for
the solution in the form
\begin{equation}
   \epsilon(\tau,p) = \epsilon_0(\tau,p) + \epsilon_1(\tau,p) +
      \epsilon_2(\tau,p) + \cdots
   \label{series}
\end{equation}
where $\epsilon_0(\tau,p)$ is the energy distribution
which starts the iteration. It is related to the number of hard gluons
by
\begin{equation}
        N_h (\tau) = \int_p^\infty  \!dk\, k^{-1}\epsilon_0(\tau, k)
      \label{NHard}
\end{equation}
so long as $p \gg T$ and $p \ll Q_s$.
Due to the one-dimensional expansion it decreases
  as \mbox{\boldmath$1/\tau$}.  Solving for
$\epsilon_1(\tau,p)$, we find, for $p\ll Q_s$,
\begin{equation}
   \epsilon_1(\tau, p) = {\epsilon_1(\tau)\over p^{1/2}}
   \label{eps1}
\end{equation}
where $\epsilon_1(\tau)$ satisfies the equation
\begin{equation}
   {\d\over\d \tau}(\tau\epsilon_1(\tau)) = \alpha^2 h_0 N^{1/2}
   N_h\tau \, .
   \label{eps1dot}
\end{equation}
We see that $\epsilon_1(\tau,p)$ is singular at small $p$, and hence
at very small $p$, $\epsilon_1\gg\epsilon_0$.  One could expect that at
these small $p$'s the series (\ref{series}) is ill behaved.  However,
if one tries to do the second iteration and inserts (\ref{eps1}) into
the collision integral of Eq.\ (\ref{boltzeq}), the result is zero.
This is because $\epsilon(p)={\rm const}/p^{1/2}$ is a formal static
solution to the Boltzmann equation (\ref{boltzeq}).  This solution
describes a stationary state with a constant energy flow from high to
low momenta.  As a consequence, at small $p$ the iteration procedure
stops at the first iteration, and the solution to the Boltzmann
equation is simply $\epsilon_0+\epsilon_1$.  Notice that Eq.\
(\ref{eps1}) has the form expected in Eq.\ (\ref{esmallp}).

To solve Eqs.\ (\ref{eps1dot}) and (\ref{Tt}) one needs to know only
the total number of the hard gluons, $N_h$, but not the whole
distribution function.  This number can be parametrized by $Q_s$ and a
dimensionless constant $c$ \cite{elastic}:
\begin{equation}
   N_h = {N_c^2-1\over 4\pi^2N_c}c {Q_s^3\over\alpha (Q_s\tau)} \, .
\end{equation}
Eqs.\ (\ref{eps1dot}) and (\ref{Tt}) then can be solved.  For the
temperature, we have $T=c_T \alpha^3 Q_s^2 \tau$, where
\begin{equation}
    c_T = {3\over160\pi^2}{N_c^2-1\over N_c}{g_N\over g_E} bh_0^2c =
        {3b\over8\pi^5} N_c^3 c \ln{\<k_t^2\>\over\mD^2} \, .
   \label{T}
\end{equation}
Eq.\ (\ref{T}) is valid only with logarithmic accuracy.  This is due
to the fact that the rate of branching by a hard particle in a thermal
medium $d^2I/dxdt$ is known only with this accuracy.  Since
$k_t^4\sim\alpha^2 N_s k_{\rm br}$,
$k_t^2/\mD^2\sim\alpha^5(Q_s\tau)^2$, and the argument of the
logarithm in Eq.\ (\ref{T}) is parametrically large for
$Q_s\tau\gg\alpha^{-5/2}$.  An improved calculation of $d^2I/dxdt$ would
give a more accurate estimate for $T$
by precisely determining the constant in the logarithm in Eq.\ (\ref{T}).
Thus we arrive at the same results, with a more accurate determination
of $T$, as in part A of this section.

The authors are indebted to M.~Gyulassy, L.~McLerran, and
R.~Venugopalan for discussions.  DTS thanks RIKEN, Brookhaven National
Laboratory, and U.S.\ Department of Energy [DE-AC02-98CH10886] for
providing the facilities essential for the completion of this work.
RB acknowledges support, in part, by DFG, contract Ka 1198/4-1.
The work of AHM is supported, in part, by a DOE Grant.
The work of DTS is supported, in part, by
a DOE OJI Award.

\end{document}